\documentclass[a4paper,12pt]{article}
\usepackage{amssymb,amsmath,amsthm,mathrsfs,bm,mathtools}
\usepackage{cite,caption}
\usepackage{hyperref,authblk}
\usepackage{tikz,tikz-cd,pgfplots}
\usetikzlibrary{arrows,positioning,calc,fadings,decorations.pathreplacing,decorations}
\newcommand\Q{{\mathcal{Q}}}
\def\Qp{\mathbf{Q}_p}

\def\R{\mathbf{R}}
\def\padic{$p$-adic }

\def\I{\mathcal{I}}
\def\H{\mathcal{H}_{[\kappa]}}
\renewcommand\k{\mathbf{k}}
\renewcommand\O{\mathbf{O}}
\newcommand\p{\mathbf{P}}
\newcommand\F{\mathbf{F}}
\newcommand\kstar{\mathbf{k}^{\star}}
\newcommand\kext{{\mathbf{k}^{\star}_{\kappa}}}

\newcommand\eq[1]{(\ref{#1})}
\newcommand\C{\mathscr{C}}

\newcommand\rt{\longrightarrow}
\newcommand\map{\longmapsto}
\newcommand\ad{\operatorname{Ad}}
\newcommand\ads{\ensuremath{{AdS}(2,\k)}}
\newcommand\lbar{{[\kappa]}}
\newcommand\abar{{[\sigma]}}
\newcommand\ksq{\kstar/(\kstar)^2}
\newcommand\stab[1]{\operatorname{Stab}{#1}}
\def\pa{\partial}
\title{Holography on local fields via Radon Transform}
\author{Samrat Bhowmick\thanks{email: tpsb5@iacs.res.in}~}
\author{Koushik Ray\thanks{email: koushik@iacs.res.in}}
\affil{\normalsize Department of Theoretical Physics, \authorcr Indian
Association for the Cultivation of Science,\authorcr Kolkata 700 032. India.}
\begin{document}
\maketitle
\thispagestyle{empty}
\begin{abstract}
\noindent
We define Radon transform and its inverse on the two-dimensional 
anti-de Sitter space over local fields using a novel construction through a quadratic equation over the local field.  
We show that the holographic bulk reconstruction of quantum fields in
this space can be formulated as the inverse Radon transform, generalizing the
case over the reals, studied earlier.  
\end{abstract}

\newpage

\section{Introduction}
Local fields with non-Archimedean valuations extend the 
fields of real and complex numbers, which are encountered more commonly in
Physics. A variety of physical theories have been generalized to local fields, starting with the observation that the Veneziano amplitude of open bosonic strings is amenable to generalizations as integrals over $p$-adic numbers \cite{Vol87}. Such generalizations received attention 
\cite{Brekke:1993gf} primarily because they often bring out the intricate consistency of physical theories, going beyond their original 
\emph{raison d'{\^e}tre}. 
Various aspects of holography in the anti-de Sitter space over the field of $p$-adic numbers have been studied recently
\cite{Gubser:2016guj,Gubser:2017tsi,Gubser:2017pyx,BHLL,DGL}. 

In the present article, we treat the bulk-reconstruction as the
inverse Radon transform from the appropriate boundary to the bulk of the
``local" version of the two-dimensional anti-de Sitter space which we shall
denote as \ads, $\k$ denoting a local field.  
It is given, in analogy with its counterpart over the reals,
as the set of solutions to a quadratic equation in three variables over the
local field. This generalizes earlier work where the HKLL formula 
\cite{HKLL} on the Euclidean anti-de Sitter space was obtained as the inverse Radon transform \cite{bsr}. While the results are consistent with the
recent proposals 
\cite{Gubser:2016guj,Gubser:2017tsi,Gubser:2017pyx,BHLL,DGL}, 
a major difference of the present treatment is that in here $\ads$\ is
treated as an analytic manifold without allusion to the tree structure. 
In this sense, it generalizes the construction of the two-dimensional anti-de
Sitter space as a group manifold \cite{wy} and Radon transform on it. 

Radon transform on local fields has been defined and studied earlier 
\cite{chernov,kochu}. Inverse Radon transform has also been defined on local fields \cite{wang}. Indeed, the inverse of the discrete Radon transform has been used effectively in image processing, 
elevating it from being exotic to useful \cite{colonna}. 
These studies dealt with affine
spaces only. In the case at hand, we have to deal with defining the Radon
transform on \ads, in which case the affine variables satisfy a quadratic
equation. Using a recent construction of the \ads\ over a local field 
\cite{guil} we seek to generalize the formulation over the reals \cite{bsr}. 
This mimics the procedure followed in formulating the Radon transform and its inverse from the Euclidean affine space to Lobachevskian space \cite{GG}.

The advantage of this formulation is firstly in its parallelity with the
case of the real field \cite{bsr}, which is indeed contained as a 
special case in here. Moreover, its connection to the group-theoretic 
structure makes it quite geometric. Inversion of Radon transform is a
classical ill-posed problem in Mathematics in that, it requires dealing
with regularized integrals. 
Relating the boundary to bulk map to Radon transform also justifies the appearance of zeta functions in the formulas \cite{Gubser:2017pyx}. 
Finally, while the result agrees with the previous proposals, 
it clarifies certain aspects of the choice of bulk and
boundary variables and demonstrates the difficulty in extending the procedure
to higher dimensions. We also note that, like its real counterpart, 
the fields are required to have a particular scaling dimension on the null cone. Let us stress that the results obtained here are valid
for all locally compact fields, including the field of $p$-adic numbers,
$\Qp$, and its ramifications, when $p$ is an odd prime. 
The explicit formulas are written for $\Qp$
to simplify notation; the generalization to other fields being
straightforward, if cumbersome.

\ads\ is looked upon as the set of solutions of  a quadratic equation in the
three-dimensional affine space $\k^3$.
Solutions of the equation fall on the orbit of the orthogonal group 
preserving the
quadratic form. Analyzing the orbits using the transitive action of the
orthogonal group it is found that the orbit splits into two disjoint lobes
separated by the null cone, the latter
given by the zeroes of the quadratic form. One
of the lobes, as for the reals, is chosen to be \ads. The
Radon transform of functions on the \ads\ is 
defined by restricting the functions on the analogue of a  two-dimensional 
plane intersecting the null cone.
It is then shown, again parallel to the reals \cite{bsr}, that if
the restricted functions scale with a specific exponent, alias, scaling
dimension, on the null cone, then the inverse Radon transform provides the
kernel for obtaining functions in the bulk \ads\ from the boundary. 

In the next section, we recall the aspects of a recent construction of the
upper-half plane over local fields using a quadratic equation \cite{guil}. 
In section~\ref{sec:para} we use the transitivity of the group preserving the associated quadratic form to parametrize the plane. In section~\ref{sec:radon} 
we define the Radon transform and its inverse and derive the boundary-to-bulk kernel as the inverse transform. The bulk reconstruction formula in terms of a kernel acting on boundary fields is obtained in section~\ref{sec:recon}.
Some computations are postponed to the
Appendix.

\section{Construction of \ads}
In this section, we review a construction of the two-dimensional anti-de Sitter
space over a locally compact field, namely, \ads\ as the set of solutions of a quadratic equation using the 
classification of orbits of the group of special orthogonal transformations
preserving the corresponding quadratic form \cite{guil}. The group theoretic
considerations are then used to parametrize \ads.
We shall restrict to the
case of \padic fields later, but the discussion turns out to be easier for general local fields. 

Let $\k$ denote a locally compact non-discrete field
with characteristic different from $2$. 
The only such fields
that are also connected are the fields of real and complex numbers. If the characteristic of $\k$ is
zero, then it is a finite extension of the field $\Qp$ of \padic numbers, for
a prime, $p$,  $p\neq 2$. If it has a non-zero characteristic $q$, 
then $\k$ is a finite extension of the field of power series over the 
residue class field modulo $q$.
Let $\kstar=\k\setminus\{0\}$ denote the
multiplicative group of $\k$. The set of squares in $\kstar$ is denoted
$(\kstar)^2$. The absolute value
 of an element $x$ of $\k$ is denoted $|x|$. The absolute value
$|x|=0$ if and only
if $x=0$. It also possesses the properties that 
$|xy|=|x||y|$ and $|x+y|\leqslant\max\big(|x|,|y|\big)$, for 
both $x$ and $y$ in $\k$. The norm of a vector 
$x=(x_1,x_2,\cdots, x_n)\in\k^n$ is defined to be 
$\parallel x\parallel = \max\big(|x_1|,|x_2|,\cdots,|x_n|\big)$. The 
integers of $\k$ is the set $\O$ of elements of $\k$ for which $|x|\leqslant
1$. We denote by $\p$ the prime ideal of elements $x$ of $\O$ such that $|x|<1$.
The residue class field is $\F=\O/\p$. The prime ideal $\p$ is principal,
that is, $\p$ contains an element $\varpi$ such that $\p=\varpi\O$. The
absolute value
of $\varpi$ is $|\varpi|=q^{-1}$, where $q$ is the order of the residue class
field $\F$. The multiplicative group $\kstar$ contains an element
$\varepsilon$ such that $\varepsilon^{q-1}=1$, implying $|\varepsilon|=1$.
Hence, $\varepsilon\in\F$. A complete set of representatives of $\F$ is given
by
$\{0,\varepsilon,\varepsilon^2,\cdots,\varepsilon^{q-2},\varepsilon^{q-1}=1\}$.

Every element of $x$ of $\k$ can be uniquely expressed as 
a convergent series
\begin{equation}
x = \varpi^n\varepsilon^k(1+a_1\varpi+a_2\varpi^2+\cdots),
\end{equation} 
where $n,k$ are real integers,
and each $a_i$ assumes value in $\F$. The elements $\varpi^n$ form
an infinite cyclic subgroup $Z$ of $\kstar$ and the 
elements $\varepsilon^k$ form a finite subgroup $Z_{q-1}$ of order $q-1$. 
The rest,
$(1+a_1\varpi+a_2\varpi^2+\cdots)$, form a compact subgroup of $\kstar$, which
we denote by $A$. For $x\in A$, $|x-1|<1$.  
Thus the multiplicative group becomes a direct product,
$\kstar=Z\times Z_{q-1}\times A$. 
We shall have to deal with the quotient group $\ksq$. It is  obtained as 
$\kstar/\sim$ upon quotienting
$\kstar$ by the equivalence relation given by  $x\sim y$ if and only if 
$y=x\ell^2$ for some $\ell\in\kstar$. The equivalence 
class of an element $x$ of $\kstar$ in $\ksq$ is written as
\begin{equation}
[x] = \{x\ell^2|x\in\kstar,\, \forall \ell\in\kstar\}.
\end{equation} 
The order of $\ksq$ is equal to the
product of the orders of the quotients
$Z/Z^2$, $Z_{q-1}/Z_{q-1}^2$ and $A/A^2$. Since
$Z=\{\varpi^n|n=1,2,3,\cdots\}$ the order of $Z/Z^2$ in $\ksq$
is $2$. Assuming $q$ to
be odd, which also guarantees that $\varepsilon$ is a non quadratic residue,
\emph{viz.} $\varepsilon\notin(\kstar)^2$, 
$Z_{q-1}$ is cyclic of even order, so that the order of 
$Z_{q-1}/Z_{q-1}^2$ in $\ksq$ is also $2$. The form of the squares of
$(1+a_1\varpi+a_2\varpi^2+\cdots)$ remain the same, so that $A^2=A$. Hence
the order of $\ksq$ is $4$ \cite{gelfand6}. 
It is partitioned into four equivalence classes,
\begin{equation}
\label{ord:k2}
\ksq = \{[1], [\varpi], [\varepsilon], 
[\varpi\varepsilon]\}.
\end{equation} 

We shall describe the construction of the anti-de Sitter space 
\ads\ given in terms of a quadratic form over $\k$. 
Let us consider
symmetric $2\times 2$ matrices 
\begin{equation}
S_2=\begin{pmatrix}
X_0 & X_1\\X_1 & X_2
\end{pmatrix}
\end{equation} 
over $\k$, transforming under $GL(2,\k)$ as
$S_2 \map {g}^TS_2 g$,
where 
\begin{equation}
\label{gl2}
g=\begin{pmatrix}
a & b\\c&d
\end{pmatrix}
\in GL(2,\k),
\end{equation} and ${g}^T$ denotes the transpose of $g$.
The determinant of $S_2$ defines a quadratic form on $\k^3$
\begin{equation}
\label{qx}
Q(X) = X_0X_2-X_1^2=\tfrac{1}{2}{X}^T\Q X,
\end{equation} 
where we define
\begin{equation}
X=\begin{pmatrix}X_0\\X_1\\X_2\end{pmatrix}\in\k^3,\quad
\Q=\begin{pmatrix}
0&0&1\\0&-2&0\\1&0&0
\end{pmatrix}.
\end{equation} 
We shall consider the set of solutions to 
\begin{equation}
\label{Ql}
Q(X)=\kappa,\quad\kappa\in\ksq, \quad -1\notin[\kappa].
\end{equation}
From \eq{ord:k2} we learn that  it suffices to consider $\kappa$ as any of 
the numbers
$1,\varpi,\varepsilon,\varpi\varepsilon$, up to homotheties.
The set of isotropic vectors $\xi\in\k^3$, namely those satisfying
$Q(\xi)=0$, is the null cone $\C$ of $Q$.
If $\k=\R$, the reals, then for $\kappa<0$ the set of solutions to
\eq{Ql} is a connected
space, a single sheet. For $\kappa>0$, on the other hand,
the set of solutions is the union of
two disconnected lobes, as shown in Figure~\ref{fig:ads}. The upper lobe is
identified as the Euclidean anti-de Sitter space.  
\begin{figure}[h]
\centering
\begin{tikzpicture}
\begin{axis}[axis lines = none, ticks=none,
    domain=0:6,
    samples=25]
\addplot3 [surf,z buffer=sort,opacity=.4,y domain=0:1.3*pi] 
    ({x*cos(deg(y))},
     {x*sin(deg(y))},  
     {x});                              
\addplot3[surf,domain=0:6,y 
domain=0:2*pi]({.7*x*cos(deg(y))},{.7*x*sin(deg(y))},{sqrt(x^2+1)});
\addplot3[surf,domain=0:6,y 
domain=0:2*pi]({-.7*x*cos(deg(y))},{-.7*x*sin(deg(y))},{-sqrt(x^2+1)});
\addplot3 [surf,z buffer=sort,opacity=.4,y domain=1.3*pi:2*pi] 
    ({x*cos(deg(y))},
     {x*sin(deg(y))},
     {x});                              
\addplot3 [surf,z buffer=sort,opacity=.4, y domain=0:1.2*pi] 
    ({-x*cos(deg(y))},
     {-x*sin(deg(y))},
     {-x});                            
\node at (rel axis cs:0.96,0.64,0.80) {$\scriptstyle\C$};
\end{axis}
\end{tikzpicture}
\caption{Two disjoint lobes of \ads\ for $\k=\R$}
\label{fig:ads}
\end{figure}
We consider a generalization of this to any local field $\k$ to
construct \ads. The condition $-1\notin[\kappa]$ in \eq{Ql} generalizes 
$\kappa>0$ over the reals.  
\ads\ is obtained by considering the orbit of a
vector in $\k^3$ which solves \eq{Ql} under the group of transformations preserving the quadratic form $Q(x)$. 
It is to be proved that the solution to \eq{Ql}
breaks into two disconnected subsets trapped within the appropriate components of the null cone and then define \ads\ from among
them, through a choice of an appropriate ``positive" component. 
 
The group 
of transformations preserving the quadratic form, $Q$ is denoted $SO(Q)$.
By \eq{qx} it is the group of matrices $M$ in $SL(3,\k)$ 
satisfying ${M}^T\Q M=\Q$.
An element \eq{gl2} of $GL(2,\k)$ corresponds 
to the linear transformation 
\begin{equation} 
X\map GX,\quad X\in\k^3,
\end{equation}
where 
\begin{equation}
\label{adG}
G = \begin{pmatrix}
a^2 & 2 ab & b^2\\ac & ad+bc&bd\\c^2&2cd&d^2
\end{pmatrix}.
\end{equation}   
Since $\det G = (\det g)^2=(ad-bc)^2$, the map  $g\map G$ defines the adjoint
representation
\begin{equation}
\ad : PGL(2,\k)\rt SL(3,\k),
\end{equation} 
which is isomorphic to the special orthogonal group $SO(Q)$ 
keeping the quadratic form $Q$ invariant.

We have the following exact sequences \cite{wikip}
\begin{equation}
\begin{tikzcd}
&0\arrow{d}&0\arrow{d}&0\arrow{d}&\\
0\arrow{r}&SZ(\k^2)\arrow{r}{zI}\arrow{d}&\kstar\arrow{r}{\det\cong
z^2}\arrow{d}&(\kstar)^2\arrow{r}\arrow{d}&0\\
0\arrow{r}&SL(2,\k)\arrow{r}\arrow{d}&GL(2,\k)\arrow{r}{\det}\arrow{d}&
\kstar\arrow{r}\arrow{d}&0\\
0\arrow{r}&PSL(2,\k)\arrow{r}\arrow{d}&PGL(2,\k)\arrow{r}\arrow{d}&\ksq\arrow{r}\arrow{d}&0\\
&0&0&0&
\end{tikzcd}
\end{equation} 
In the first row, $zI$ denotes the set of scalar transformations and $I$
the identity matrix. The last row, under the action of the adjoint
then yields the isomorphism
\begin{equation}
\label{SO:Ad}
\ad(PGL(2,\k))/\ad(PSL(2,\k)\cong
SO(Q)/\ad(PSL(2,\k))\cong \ksq.
\end{equation} 
An element of $SO(Q)$ can be decomposed as $N^+N^-H$, where
\begin{equation}
\label{Iwa}
N^+(x) = \begin{pmatrix}
1& 2x & x^2\\0&1&x\\0&0&1
\end{pmatrix},
\quad
N^-(y) = \begin{pmatrix}
1&0&0\\y&1&0\\y^2 & 2y&1
\end{pmatrix},
\quad
H(z) = \begin{pmatrix}
z&0&0\\0&1&0\\0&0&1/z
\end{pmatrix},
\end{equation} 
with $x,y\in\k$ and $z\in\kstar$ \cite{guil}. 
Each of
these is of the form \eq{adG}.

Next, we study the null cone
$\C$. It breaks into a disjoint union
of semi-cones, one of which is to be identified as the positive 
one. Let us consider  a non-zero vector on the null cone,
$\xi\in\C\setminus\{0\}$. 
Since $\xi_0\xi_2=\xi_1^2$, if $\xi_1=0$, then either 
$\xi_0\neq 0$ or $\xi_2\neq 0$. If, on
the other hand, $\xi_1^2\neq 0$, then $\xi_0\xi_2\in(\kstar)^2$. 
Since the product
of two numbers is a square, they are equal, modulo a square. Hence, there
exists  $\sigma\in\kstar$, such that both $\xi_0$ and $\xi_2$ belong to $\abar$.
This motivates the definition of semi-cones corresponding to each class
$\abar\in\ksq$ as
\begin{equation}
\label{Ca}
\C_{\abar} =\{\xi\in\C\setminus\{0\}|\xi_0, \xi_2\in\abar\cup\{0\} \}.
\end{equation} 
Thus any point on the null cone belongs to one and only one of the semi-cones,
$\xi\in\C_{\abar}$.
There are four semi-cones corresponding to the classes in \eq{ord:k2}.

Let $\stab{\C_{\abar}}$ denote the stabilizer group of the semi-cone
$\C_{\abar}$ in $SO(Q)$.
There is an isomorphism
\begin{equation}
SO(Q)/\stab{\C_{[1]}}\cong\ksq.
\end{equation} 
To see this, we first show that $N^{\pm}$ stabilizes each semi-cone
$\C_{\abar}$. Let
$\xi=\left(\begin{smallmatrix}\xi_0\\\xi_1\\\xi_2\end{smallmatrix}\right)
\in\C\setminus\{0\}$. In accordance with \eq{Ca}, $\xi\in\C_{[\xi_2]}$.
It transforms under $N^+(x)$ to another non-zero
point on the null cone,
\begin{equation}
\xi'= 
\begin{pmatrix}\xi'_0\\\xi'_1\\\xi'_2\end{pmatrix} = N^+(x)\xi = 
\begin{pmatrix}
\xi_0+2x\xi_1+x^2\xi_2\\\xi_1+x\xi_2\\\xi_2
\end{pmatrix}
\end{equation}
satisfying $Q(\xi')=0$. Three cases arise. If none of the components of $\xi$
vanishes, then since $\xi_2'=\xi_2$, it follows from \eq{Ca} that
$\xi'\in\C_{[\xi_2]}$.
Secondly, if $\xi_0=0$, $\xi_2\neq 0$ and $\xi_1=0$, then 
\begin{equation} 
\xi'= \begin{pmatrix}
x^2\xi_2\\x\xi_2\\\xi_2
\end{pmatrix}.
\end{equation}
Clearly, $[\xi_0']=[\xi_2']$, so that by \eq{Ca}, $\xi'\in\C_{[\xi_2]}$.
Finally, if $\xi_0\neq 0$ and $\xi_1=\xi_2=0$, then 
$\xi'=\xi\in\C_{[\xi_2]}$. Thus $N^+$ stabilizes each semi-cone.
Similar arguments hold for $N^-$. 

Let us now consider $SO(Q)/\stab{\C_{[1]}}$. Using the decomposition 
\eq{Iwa} and the fact that both $N^+$ and $N^-$ stabilize $\C_{[1]}$, we have
\begin{equation}
SO(Q)/\stab{\C_{[1]}} \cong H/\stab{\!}_H{\C_{[1]}}.
\end{equation} 
Since the action of $H$ on $\xi$ is  
\begin{equation} 
H(w)\xi=\begin{pmatrix}w\xi_0\\\xi_1\\\xi_2/w\end{pmatrix}, 
\end{equation} 
and the classes $[\xi_0]=[\xi_2]$ are defined modulo squares,
the stabilizer in $H$ of $\C_{[1]}$ is the subgroup 
\begin{equation}
\begin{pmatrix}
w^2 &0&0\\0&1&0\\0&0&1/w^2
\end{pmatrix},\quad w\in\kstar,
\end{equation} 
which is isomorphic to $(\kstar)^2$.
Hence,  
\begin{equation}
\label{SO:stab}
SO(Q)/\stab{\C_{[1]}} \cong H/\stab{\!}_H{\C_{[1]}}\cong \ksq.
\end{equation} 
Using this and the isomorphism \eq{SO:Ad} we identify
$\ad({PSL(2,\k)})$ and $\stab{\C_{[1]}}$.

We now have a decomposition of the null cone into semi-cones. We need to
understand the decomposition of the hyperboloids, obtained as solutions to
\eq{Ql}, under the group $\stab{\C_{[1]}}$. The hyperboloids are
obtained as the orbit under $SO(Q)$ of any point solving the equation. 
The point 
\begin{equation}
\label{vl}
v_0 = \begin{pmatrix}\kappa\\0\\1\end{pmatrix}
\end{equation} 
in $\k^3$ solves \eq{Ql}, hence belongs to the hyperboloid. 
We study the orbits $SO(Q)\cdot{v_0}$. Using the isomorphism 
$\ad({PSL(2,\k)})\cong\stab{\C_{[1]}}$, we  classify these orbits in
terms of the orbits  $\ad({PSL(2,\k)})\cdot{v_0}$.

Since $-1$ does not belong to the
class $\lbar$ by assumption, the orbit $SO(Q)\cdot v_0$ 
decomposes into two disjoint subsets under
$PSL(2,\k)$. These are the analogues of
the two lobes of \eq{Ql} for $\kappa>0$ on $\k=\R$, as in 
Figure~\ref{fig:ads}. The
proof requires understanding the stabilizer of $v_0$. 
The stabilizer of $v_0$ fixes the plane orthogonal to $v_0$ too. A
non-zero vector $v_{\perp}\in\k^3$ is orthogonal to $v_0$ if ${v_{\perp}}^T\Q v_0=0$. 
The most general such vector is 
\begin{equation} 
\label{vp}
v_{\perp}=\begin{pmatrix}
-\kappa a\\b\\a
\end{pmatrix}, \quad a, b\in\k.
\end{equation} 
These can be verified by writing an element of $\stab{v_0}$ in $SO(Q)$
in the parametric form
\begin{equation}
\begin{split}
\label{S0}
S_0&=\frac{1}{1+\kappa\theta^2}
\begin{pmatrix}
1 & -2\kappa\theta & \kappa^2\theta^2 \\
\theta & 1-\kappa\theta^2 & -\kappa\theta\\
\theta^2 & 2\theta & 1
\end{pmatrix}\\
&=
N^-(\theta)H\big(\tfrac{1}{\kappa\theta^2}\big)N^+(-\kappa\theta)
\end{split}
\end{equation} 
where $\theta\in\k$.
If $v_{\perp}$ intersects the null cone, 
then $Q(v_{\perp})=-\kappa a^2-b^2 =0$, which in turn signifies 
$\kappa=-b^2/a^2$. But since $-1$ does
not belong to $\lbar$, this is impossible. Hence, the vector 
$v_{\perp}$ does not
intersect the null cone. In order to obtain a plane stabilized by
$\stab{v_0}$ which intersects the null cone we shift the origin of the
orthogonal plane and consider the affine plane $\Sigma$ of 
vectors $\tfrac{1}{\kappa} v_0+v_{\perp}$. A vector $X\in\k^3$ is 
in $\Sigma$ provided $X-\tfrac{1}{\kappa}v_0$ is in the form \eq{vp}, implying
\begin{equation}
X_0-1=-\kappa a,\quad X_2-\tfrac{1}{\kappa} = a.
\end{equation} 
Eliminating $a$ relates the components of $X$.
Thus $X\in\Sigma\cap\C$ if and only if
\begin{gather}
\label{int1}
\kappa X_2+X_0=2\\
\label{int2}
X_0X_2=X_1^2.
\end{gather}
Transitivity of the action of $\stab{v_0}$ on $\Sigma\cap\C$ guarantees the
existence of a class
$\abar\in\ksq$ such that $\Sigma\cap\C_{\abar}\neq\emptyset$. 
The second condition \eq{int2} implies, as in \eq{Ca}, that there is 
a $\sigma$ in $\kstar$
such that $X_0,X_2\in\abar$. Then \eq{int1} implies $1\in\abar+\kappa\abar$.
Since $-1\notin\lbar$ and $p\neq 2$, in case we choose $\k=\Qp$ 
(else the right side of \eq{int1} vanishes modulo $p$), this implies, 
in turn, that $1\in\abar$ or
$1\in\kappa\abar$. Thus the set of $\sigma$ for which $\Sigma$ intersects 
$\C_{\abar}$ is 
\begin{equation} 
\{\kappa+x^2 \in\ksq| x\in\kstar\}.
\end{equation} 
The lift of this in $\kstar$, namely,
\begin{equation}
\label{kl}
\kext = \{x^2+\kappa y^2| x,y\in\kstar\},
\end{equation} 
is called the norm subgroup of $\kstar$. The 
index of $\kext$ in $\ksq$ is $2$ \cite{guil}. Since $\stab{v_0}$ permutes the
semi cones $\C_{\abar}$ which intersect $\Sigma$, it has two distinct orbits
among the set of $\C_{\abar}$. Thus the orbit of $v_0$ is a two-sheeted
hyperboloid. 

We now need to choose one of the sheets, preferably the positive one,
analogous to the upper lobe in Figure~\ref{fig:ads} in 
the real case. Positivity in the case at hand is determined by declaring the
elements of the norm subgroup \eq{kl} to be positive. The solution to \eq{Ql}
decompose into two disjoint orbits under $\kext\ad{PSL(2,\k)}$, provided
$-1\notin\kext$. Starting now with $v_0$, it can be proved \cite{guil}
that the set $\kext\ad{PSL(2,\k)}$ and the set 
\begin{equation}
\label{halpha}
\H=\{X\in\k^3|\forall\xi\in\C_{[1]}, B(X,\xi)\in\kext\}
\end{equation}  
are equal. Here $B(x,y)$ denotes the polar bilinear form defined as
\begin{equation}
\label{polar}
B(x,y) = \tfrac{1}{2}\big(Q(X+Y)-Q(X-Y)\big),\quad X,Y\in\k^3.
\end{equation} 
Finally. the local version of the anti-de Sitter space, \ads, is then given
as the ``upper" subset of $\H$,
\begin{equation}
\label{ads:def}
\text{\ads}=\{X\in\k^3|\forall\xi\in\C_{[1]},\ 
Q(X)=\kappa, B(X,\xi)\in\kext\}.
\end{equation}  
\subsection{Parametrization}
\label{sec:para}
Transitivity of the action of $SO(Q)$ on \ads\ allows writing any point on the
latter in a parametric form. A generic point on \ads\ 
is obtained from $v_0$ as $g_0v_0$, 
\begin{equation}
\label{para:x}
\begin{pmatrix}
X_0\\X_1\\X_2
\end{pmatrix} = g_0v_0 = 
\begin{pmatrix}
(\kappa z^2+x^2)/{z}\\
x/{z}\\
1/{z}
\end{pmatrix},
\end{equation} 
where $g_0=N^+(x)H(z)$, $x\in\k$, $z\in\kstar$.
The coordinates of the semi-cone $\C_{[1]}$ are parametrized as 
\begin{equation}
\label{para:xi}
\begin{split}
\xi_0 &= \eta^2\tilde{x}^2\\
\xi_1 &= \eta^2 \tilde{x}\\
\xi_2 &= \eta^2.
\end{split}
\end{equation} 
Using these parametrizations we have, by \eq{polar}
\begin{equation}
\label{b:xix}
B(X,\xi) = \frac{\eta^2}{z} (\kappa z^2+(x-\tilde{x})^2).
\end{equation} 
In order for this to be in the norm subgroup
$\kext$, the variable  $z$ must be a square,
\begin{equation}
\label{zw2}
z=w^2,\quad w\in\kstar.
\end{equation} 
Given a point $\xi$ on the null cone $\C_{[1]}$, written as \eq{para:xi},
we shall consider the family  of curves in $\H$, given by 
\begin{equation}
\label{sigma}
\Sigma_s:=\xi\cdot X+s=0,
\end{equation} 
where $s\in\k$ is a parameter and 
\begin{equation}
\label{xix}
\xi\cdot X={\xi}^T\mathcal{Q}X 
= \tfrac{1}{2}B(X,\xi),
\end{equation} 
where we used \eq{para:x} and \eq{para:xi}. With these parametrization of
\ads\ and $\C_{[1]}$ we now proceed to define the Radon transform.
\section{Radon Transform in \ads\ and its inverse}
\label{sec:radon}
Radon transform \cite{kochu} and its inverse \cite{wang} have been studied
in affine spaces over local fields, in particular, on $\Qp^n$. 
In this section we generalize these to \ads\ 
following the generalization of the 
the transforms from the affine Euclidean space to the Euclidean 
anti-de Sitter space \cite{GG},
used earlier to obtain the bulk reconstruction formulas \cite{bsr}.
Indeed, the expressions for the Radon transform and its inverse over
the latter carry over, \emph{mutatis mutandis}, to \ads. However, for 
concreteness we shall assume, from now on, $\k=\Qp$. This will allow us 
to write explicit formulas like \eq{c:val}.

For a smooth function $f$ on \ads\ we define its Radon transform by 
restricting it on a curve
$\Sigma_{\kappa}$, given by \eq{sigma},
\begin{equation}
\label{Radon}
\check{f}(\xi) = \int\displaylimits_{Q(X)=\kappa} 
f(X) \delta(\xi\cdot X - \kappa) d^3X,
\end{equation}
where $d^3X$ denotes the measure on $\k^3$, $d^3X=dX_0dX_1dX_2$, where restriction on the surface $Q(X)=\kappa$ is given by the Gel'fand-Leray
form 
\begin{equation}
\label{X:restrict}
\begin{split}
\int\displaylimits_{Q(X)=\kappa} d^3X &= \int\limits_{Q(X)=\kappa}
\frac{dX_1dX_2}{\left|\frac{\pa Q(X)}{\pa X_0}\right|} \\
&= \int\limits\frac{dx\ dz}{|z|^2},
\end{split}
\end{equation} 
using the parametrized expression \eq{para:x}.
The measure is invariant under $SO(Q)$ transformations. In terms of this
parameterization
\begin{equation}
\xi\cdot X-\kappa=
\frac{\eta^2}{2z}\left(\kappa z^2+(x-\tilde{x})^2\right)-\kappa.
\end{equation}
The holographic boundary is located at $x=\tilde{x}$, as $z\rt 0$ on $\Sigma_\kappa$.
The inverse Radon transform is given as the integral of $\check{f}$ over the
null cone,
\begin{equation}
\label{invRadon}
f(X) = c\int\displaylimits_{\substack{Q(\xi)=0}}
\frac{\check{f}(\xi)}{|\xi\cdot X - \kappa|^2} d^3\xi, 
\end{equation}
exactly as in the real case \cite{bsr}, with the modulus
changed to the valuation of $\k$ in the case at hand,
with $Q(X)=\kappa$ and 
\begin{equation}
\label{c:val}
c = \left|\frac{\sqrt{\kappa}}{2}\right|\frac{1}{\max\big(1,|\kappa|^{-2}\big)}
\frac{\zeta(2)\zeta(1)}{\zeta(0)
\big(1-2\zeta(1)\big)},
\end{equation} 
where $\zeta(x)$ denotes the $p$-adic zeta function.

Let us prove that \eq{invRadon} is the inverse of \eq{Radon}.
For this to hold, the former plugged in the latter should give back $f(X)$, 
that is,
\begin{equation}
\label{invRadc}
\begin{split}
f(X)&=c\int\displaylimits_{\substack{Q(\xi)=0\\Q(X')=Q(X)=\kappa}} 
f(X')\frac{\delta(\xi\cdot X' - \kappa)}{|\xi\cdot X - \kappa|^2} d^3\xi d^3X', 
\end{split}
\end{equation}
or, equivalently,
\begin{equation}
\label{dirac}
c\int\limits_{Q(\xi)=0}\frac{\delta(\xi\cdot X' - \kappa)}{%
|\xi\cdot X - \kappa|^2} d^3\xi 
=\delta^{(2)}(X-X'),
\end{equation}
where the Dirac distribution on the RHS is on \ads. 
Let us demonstrate this for a pair of nearby points. Following the treatment in
the case of reals, let us choose $X'=v_0$. Any other point \eq{para:x} 
on \ads\ can be transformed to 
\begin{equation}
X=\begin{pmatrix}y\\0\\\kappa/y\end{pmatrix}
\end{equation}  
by an appropriate element of $\stab{v_0}$, given in \eq{S0}. 
Let us note that $X$ goes to $X'$ as $y$ limits to $\kappa$. 
Inserting $X$ and $X'$ in the integral on the LHS of \eq{dirac} we have
\begin{equation}
\begin{split}
\mathcal{J} &=
\int\limits_{Q(\xi)=0}\frac{\delta(\xi\cdot X' - \kappa)}{%
|\xi\cdot X - \kappa|^2} d^3\xi \\
&=\int_{\k^2} \frac{\delta(\frac{\xi_0+\kappa\xi_2}{2}-\kappa)}{%
\left|\frac{\kappa\xi_0/y+y\xi_2}{2}-\kappa\right|^2 }
\frac{d\xi_0 d\xi_2}{\left|\xi_0\ \xi_2\right|^{1/2}
}%
\end{split}
\end{equation} 
By a change of variable
\begin{equation}
\hat{\xi}=\frac{\xi_0+\kappa\xi_2}{2},
\end{equation} 
such that the numerator becomes $\delta(\hat{\xi}-\kappa)$, we can perform the
integration on $\hat{\xi}$. Further defining $\sigma=\xi_2/2$, the integral
becomes 
\begin{equation}
\mathcal{J}
=\left|\frac{2}{\sqrt{\kappa}}\right| \left|\frac{y}{y+\kappa}\right|^2
\left|\frac{1}{y-\kappa}\right|^2\int_{\k}\frac{d\sigma}{%
\left|\sigma-\tfrac{\kappa}{y+\kappa}\right|^2
\left|\sigma(1-\sigma)\right|^{1/2}
}.%
\end{equation} 
Let us recall that the delta distribution in two dimensions, $\Qp^2$,  
is defined as the limit of the Riesz kernel \cite{vega}
\begin{equation}
\delta^{(2)}(x)=\lim_{a\rt 0}\frac{\zeta(2)}{\zeta(a)}\frac{1}{
\parallel x\parallel^{2-a}},\quad x\in\Qp^2.
\end{equation} 
Furthermore, 
$\parallel X-X'\parallel=|y-\kappa|\max\big(1,0,\frac{1}{|y|}\big)$. 
In order to determine the constant $c$ it suffices to work in the 
limit $y\rt\kappa$.
The integral in $\mathcal{J}$ is then evaluated in terms of 
\begin{equation}
\begin{split}
\I(1/2,-1/2,-2)
&=\frac{1-2\zeta(1)}{\zeta(1)}\\
&=-(1+1/p),
\end{split}
\end{equation} 
worked out in \eq{Iabc}. 
Plugging the expressions for $\mathcal{J}$ and $\delta^{(2)}(X-X')$ in the
left and right sides of \eq{dirac}, respectively, 
yields \eq{c:val}.
\section{Bulk reconstruction}
\label{sec:recon}
We use the inverse Radon transform for the boundary to bulk
reconstruction, following the strategy in the case of reals \cite{bsr}. 
Using the Gel'fand- Leray form of the measure on the null cone \eq{invRadon}, we obtain 
\begin{equation}
\label{bo:bu}
\begin{split}
f(X) &= c\int\displaylimits_{Q(\xi)=0}
\frac{\check{f}(\xi)}{|\xi\cdot X - \kappa|^2} d\xi_0d\xi_1d\xi_2\\
&= c\int\displaylimits_{\k^2}
\frac{\check{f}(\xi)}{|\xi\cdot X - \kappa|^2}
\frac{d\xi_1d\xi_2}{\left|\frac{\pa Q(\xi)}{\pa\xi_0}\right|}\\
&= c\int\displaylimits_{\k^2}
\frac{\check{f}(\xi)}{
\left|\frac{\eta^2}{2z}\left(\kappa z^2+(x-\tilde{x})^2\right)-\kappa\right|^2}
d\eta^2 d\tilde{x},
\end{split}
\end{equation} 
where we used the parametrization \eq{para:xi}.
We now assume that the function $\check{f}$ on the null cone has a scaling 
dimension $\Delta$, that is 
\begin{equation}
\label{scal}
\begin{split}
\check{f}(\xi)&=\check{f}(\xi_0,\xi_1,\xi_2)\\
&= \check{f}(\eta^2\tilde{x}^2,\eta^2\tilde{x},\eta^2)\\
&=\frac{1}{|\eta|^{2\Delta}}\tilde\phi(\tilde{x}),
\end{split}
\end{equation} 
where we denoted $\tilde\phi(\tilde{x})=\check{f}(\tilde{x}^2,\tilde{x},1)$. 
This is a function on the boundary.
This assumption is essential in interpreting the field $\tilde{\phi}$  
as a conformal field on the boundary.
Substituting this form and defining a new variable 
\begin{equation}
t^2 = \frac{\eta^2}{2z} \big(\kappa z^2+(x-\tilde{x})^2\big)
\end{equation} 
\eq{bo:bu} yields the bulk function
\begin{equation}
\begin{split}
\label{b-b}
\phi(z,x) &= \frac{1}{|2|^{\Delta-1}}\frac{c}{|\kappa|^{\Delta+1}} 
\left(\int_{t\in\k}\frac{dt^2}{|t^2|^{\Delta}|t^2-1|^2}\right)%
\int_{\k}\left|\frac{\kappa z^2+(x-\tilde{x})^2}{z}\right|^{\Delta-1}
\tilde\phi(\tilde x)\,d\tilde{x} \\
&= \frac{c}{2}\frac{1}{|2|^{\Delta-2}}\frac{1}{|\kappa|^{\Delta+1}} 
\mathcal{I}(2(1-\Delta),-2,-2)\int_{\k}
\left|\frac{\kappa z^2+(x-\tilde{x})^2}{z}\right|^{\Delta-1}\tilde\phi(\tilde x)\,d\tilde{x}
\end{split}
\end{equation} 
where we  use \cite{pQM}
\begin{equation}
dt^2=\tfrac{1}{2}|2t|dt
\end{equation} 
and write $\phi(z,x)$ for $f(X)$. 
Let us note that since $p>2$, $|2|=1$. The integral  
$\I$ is defined in the Appendix.  From \eq{Iabc} we derive
\begin{equation}
\I(2(1-\Delta),-2,-2) =
\frac{\zeta(2\Delta+2)-\zeta(2\Delta-1)-2\zeta(1)-2\zeta(2)+2}{\zeta(1)}.
\end{equation} 
Supposing that the above properties of functions hold good for quantum
fields, \eq{b-b} yields a scalar field $\phi(z,x)$ in the bulk of \ads\
in terms of the boundary field $\tilde\phi(\tilde x)$
though a kernel. This generalizes the formula over the reals, $\k=\R$
\cite{bsr}.
\section{Summary}
To summarize, we have considered the realization of 
the two-dimensional anti-de Sitter space over a
local field $\k$ as the set of by the solutions of
a quadratic equation \eq{Ql} over $\k$. Upon identifying the appropriate
portion of the null cone containing one of the two disjoint lobes of the
set identified as \ads, we define the Radon transform of functions in it. 
This is facilitated by the transitive action of $SO(Q)$ on \ads,
since it allows us to parametrize the latter as well as the null cone.
We also obtain the inverse of the Radon transform. Assuming that these
definitions are valid for \emph{local fields on local fields} and that the
scalar  fields have specific scaling behavior \eq{scal} on the null cone, we
derive a kernel that takes a boundary field to the AdS bulk \eq{b-b}.

The kernel \eq{b-b} is similar to the one proposed earlier 
\cite{Gubser:2017pyx}. However, as in \eq{zw2}, the coordinate $z$ is valued
not just in $\k$, but is a square.
Another important difference is that the present treatment
does not refer to the tree structure. It thus deals with analytic
expressions, and are valid for any local field. Using the identification of
the \ads\ with the hyperbolic disk, obtained by projecting it on
a sphere, \ads\ may be related to the tree \cite{guil}. This can be used
to compare the two approaches. The present treatment also clarifies
the difficulties of
extending the computations to higher dimensional anti-de Sitter spaces over
$\k$. In $n$ dimensions, the quadratic form  becomes 
$Q(X)=X_0X_{n+1}-\sum_{i=1}^n X_i^2$ and \eq{Ql} is to be accordingly
generalized. However, unless $\k=\R$, the arguments leading to the existence
of $\alpha$ in \eq{Ca} fails. Furthermore, the derivation of the norm
subgroup \eq{kl} from \eq{int1} is restricted to $p\neq 2$ 
in the instance $\k=\Qp$.
It will be interesting to see how going over to the tree may 
avoid these difficulties, if at all.  The present treatment, however,  is
in exact analogy with the bulk reconstruction over the reals \cite{bsr}.
The identification of the boundary in the present approach is, in our opinion,
more intuitive. We hope that this approach will be useful in carrying out
analytic computations of holographic duality over local fields. 
\appendix 
\renewcommand{\theequation}{\thesection.\arabic{equation}}
\setcounter{equation}{0}
\section{Evaluation of $\I(\alpha,\beta,\gamma)$}
In this appendix we evaluate the various integrals written in the 
text. Here $\k=\Qp$.
We have \cite{VVZ,vlad}, for $x\in\Qp$,
\begin{equation}
\label{int:1}
\begin{split}
\int_{|x|<1} |x|^{\delta-1} dx
&= \int_{|x|\leqslant 1} |x|^{\delta-1} dx -
\int_{|x|=1} dx \\
&= \frac{\zeta(\delta)-1}{\zeta(1)},
\end{split}
\end{equation} 
where the $p$-adic zeta function is defined as
\begin{equation} 
\label{zeta:def}
\zeta(x) = \frac{1}{1-p^{-x}},\quad x\in\Qp.
\end{equation} 
The zeta function satisfies the relation
\begin{equation}
\label{zeta:rel}
\zeta(-\alpha)=1-\zeta(\alpha).
\end{equation} 
Any $x\in\Qp$ is written as
\begin{equation}
x=x_0+x_1p+x_2p^2+\cdots.
\end{equation}
The integral
\begin{equation}
\begin{split}
\label{Ia}
\I(\alpha)&=\int_{\substack{|x|=1\\x_0=\varepsilon}} dx\
|x-\varepsilon|^{\alpha}\\
&=\int_{x_1\neq 0} |x|^{\alpha}
+\int_{\substack{x_1=0,\\x_2\neq 0}} dx |x|^{\alpha}
+\int_{\substack{x_1=x_2=0,\\x_3\neq 0}} dx |x|^{\alpha} +\cdots\\
&=\sum_{n=1}^{\infty} p^{-n\alpha} \int_{|x|=1} dx\\
&=\frac{\zeta(\alpha)-1}{\zeta(1)},
\end{split}
\end{equation} 
where $\varepsilon$ is a non-zero element of $\mathbf{F}$.
Let 
\begin{equation}
\I(\alpha,\beta,\gamma) = \int_{\Qp} dx |x|^{\alpha-1} |x-\varepsilon_1|^{\beta}
|x-\varepsilon_2|^{\gamma},
\end{equation} 
where $\varepsilon_1\neq\varepsilon_2$ are elements of the residue class
field $\mathbf{F}$.
The integral is evaluated by breaking it into pieces.
\begin{equation}
\begin{split}
\label{Iabc}
\I(\alpha,\beta,\gamma) 
&= \int_{|x|<1} dx|x|^{\alpha-1} 
+\int_{|x|>1}|x|^{\alpha+\beta+\gamma-1} +
\int_{|x|=1}dx |x-\varepsilon_1|^{\beta} |x-\varepsilon_2|^{\gamma}\\
&= \int_{|x|<1} dx|x|^{\alpha-1} 
+ \int_{|x|<1} dx|x|^{-\alpha-\beta-\gamma-1} \\
&\qquad+\int_{\substack{|x|=1,\\x_0\neq\varepsilon_1,x_0\neq\varepsilon_2}} dx
+\int_{\substack{|x|=1,\\x_0=\varepsilon_1}} dx |x-\varepsilon_1|^{\beta}
+\int_{\substack{|x|=1,\\x_0=\varepsilon_2}} dx |x-\varepsilon_2|^{\gamma}\\
&=\frac{
\zeta(-\alpha-\beta-\gamma)-\zeta(-\alpha)-\zeta(-\beta)-\zeta(-\gamma) 
-2\zeta(1)+2}{\zeta(1)}, 
\end{split}
\end{equation} 
where we used \eq{int:1}, \eq{zeta:def}, \eq{Ia} and \eq{zeta:rel} in the final step.
Let us note that the exact values of $\varepsilon_1$ and $\varepsilon_2$ are not
important for evaluating the integral as long as they are unequal. 
\subsection*{Acknowledgement}
We thank Shamik Basu and Siddhartha Sen for very useful conversations.

\end{document}